# GRACES: Gemini remote access to CFHT ESPaDOnS Spectrograph through the longest astronomical fiber ever made (Experimental phase completed.)

André-Nicolas Chené*[a], John Pazder[b], Gregory Barrick[c], Andre Anthony[b], Tom Benedict[c], David Duncan[b], Pedro Gigoux[a], Scot Kleinman[a], Lison Malo[c], Eder Martioli[d], Claire Moutou[c], Vinicius Placco[a], Vlad Reshetovand[b], Jaehyon Rhee[e], Katherine Roth[a], Ricardo P. Schiavon[f], Eric Tollestrup[a], Tom Vermeulen[c], John White[a], Robert Wooff[b]

[a]Gemini Observatory, Northern Operations Center, 670 North A'ohoku Place, Hilo, HI 96720, U.S.A.; [b]Herzberg Institute of Astrophysics, National Research Council of Canada, Victoria, BC V9E 2E7, Canada; [c]Canada-France-Hawaii Telescope Corporation, 65-1238 Mamalahoa Hwy Kamuela, Hawaii 96743, U.S.A.; [d]Laboratório Nacional de Astrofísica (LNA/MCTI), Rua Estados Unidos, 154 Itajubá, MG, Brazil; [e]Oregon State University, Physics department, 301 Weniger Hall, Corvallis, OR 97331, U.S.A.; [f]Astrophysics Research Institute, Liverpool John Moores University, Twelve Quays House, Egerton Wharf, Birkenhead CH41 ILD, UK

## ABSTRACT

The Gemini Remote Access to CFHT ESPaDONS Spectrograph has achieved first light of its experimental phase in May 2014. It successfully collected light from the Gemini North telescope and sent it through two 270 m optical fibers to the the ESPaDOnS spectrograph at CFHT to deliver high-resolution spectroscopy across the optical region. The fibers gave an average focal ratio degradation of 14% on sky, and a maximum transmittance of 85% at 800nm. GRACES achieved delivering spectra with a resolution power of R = 40,000 and R = 66,000 between 400 and 1,000 nm. It has a ~8% throughput and is sensitive to target fainter than 21[st] mag in 1 hour. The average acquisition time of a target is around 10 min. This project is a great example of a productive collaboration between two observatories on Maunakea that was successful due to the reciprocal involvement of the Gemini, CFHT, and NRC Herzberg teams, and all the staff involved closely or indirectly.

**Keywords:** spectroscopy, high-resolution, fiber, optical, Gemini North telescope, Canada-France-Hawaii Telescope, ESPaDOnS

## 1. INTRODUCTION

Gemini Remote Access to CFHT ESPaDOnS Spectrograph (GRACES) is a collaborative instrumentation experiment between the Canada-France-Hawaii Telescope (CFHT), Gemini, and NRC-Herzberg in Canada. It combines the large collecting area of the Gemini North telescope with the high resolving power and high efficiency of the ESPaDOnS[1] spectrograph at CFHT, to deliver high-resolution spectroscopy across the optical region. This is achieved through two 270 m optical fibers (the longest ever made for astronomy) fed from the Gemini telescope to the spectrograph. The goal of the experiment is to demonstrate that a long fiber fed spectrograph can produce an efficient, sensitive instrument that delivers useful data to the Gemini community users and that gives performances comparable with other high-resolution spectrographs on 6-10m class telescopes. The final objective of the experiment is to provide in a near future the Gemini Observatory with a new capacity at a small fraction of the cost of building a whole new instrument[2].

GRACES was successfully installed and tested on-sky in April-May 2014! This paper presents our results after GRACES first installation at the telescope and first on-sky commissioning. In Section 2, we briefly describe the components and show their performances. In Section 3, we describe how GRACES operates. In Section 4, we discuss the data format and the reduction procedure. Finally, we present in Section 5 the quality of the data observed during the May 2014 on-sky commissioning and derive GRACES high-resolution spectroscopic performances, before concluding this work in Section 6.

*achene@gemini.edu; phone 1 808 974-2632; fax 1 808 974-2589; gemini.edu

## 2. GRACES COMPONENTS

GRACES consists of three components: 1) an **injection module** sending the light from the Gemini telescope into the GRACES fibers, 2) two 270m-long **GRACES fibers** and 3) a **receiver unit** responsible for injecting the light from the fibers into the ESPaDOnS spectrograph at the Canada-France-Hawaii Telescope (CFHT). A complete description of the GRACES components was already presented in previous SPIE proceedings[3]. This section describes how the components performed during the acceptance test (achieved in the NRC-Herzberg optical lab from 3 to 7 Mar, 2014) and during the on-sky commissioning (in April-May, 2014).

### 2.1 The injection module

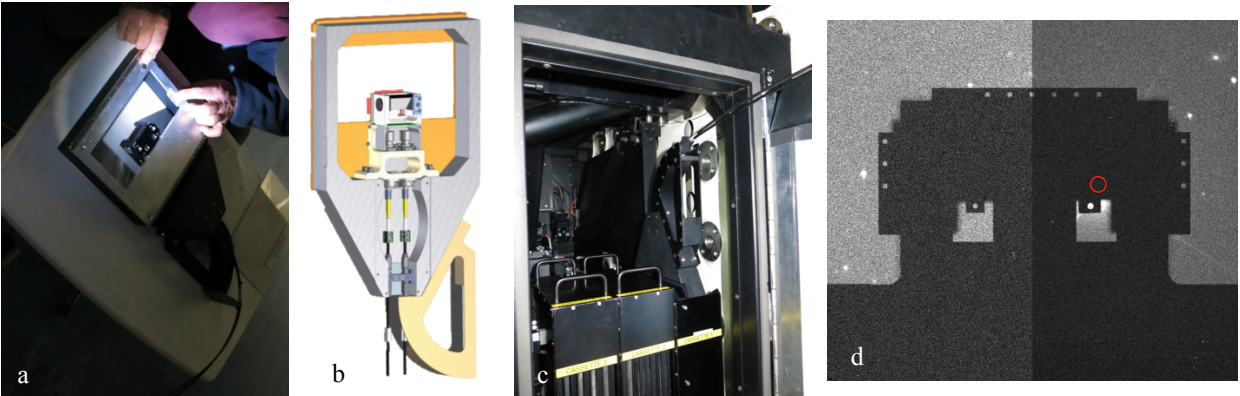

Figure 1. a) The injection module in its cassette during alignment. b) CAD model of the module showing the pickoff mirror, the injection lenses, the fibers and the cassette. c) The cassette installed into GMOS, next to the GMOS masks holders. d) Shadow image of the injection module taken with GMOS in imaging mode when a bright star was centered on one fiber (the position where the target has to be centered is marked by a red circle). We can see the silhouette of the mask containing the fiducial holes that were used to calibrate the acquisition of the target during on-sky commissioning.

The injection module consists of a pickoff mirror directing the light beam coming from the telescope into the lenses that feed the fibers (see Figure 1a and b). It is mounted on a cassette that is installed into the Gemini Multi-Object Spectrograph (GMOS)[4]. That cassette is a slightly modified version of the one that was used for the decommissioned Gemini instrument named bench-Mounted High-Resolution Spectrograph (bHROS)[5], and is installed into the rack that usually holds the GMOS Integral Field Unit (see Figure 1c). The whole module keeps its alignment within 0.05 arcsec when it is exposed to temperature variations between +10°C and −10°C. It also stays aligned when the fibers are disconnected and reconnected to the module and when the cassette is submitted to flexures. Once installed into GMOS, the cassette can be moved in and out the beam repeatedly within 0.01 arcsec. Flexures in the cassette can be as big as 0.14 arcsec at low telescope elevation. However, since a correction is applied using a model of the GMOS flexure observations on-sky, the accuracy target acquisition procedure placing the target into the GRACES science fiber (see Section 3.3) is lower than 0.07 arcsec 95% of the time, with a maximum of 0.12 arcsec at elevations lower than 30°.

### 2.2 The 270m long fibers

A complete description of the GRACES fibers (built by FiberTech Optica, http://fibertech-optica.com, in collaboration with NRC Herzberg) and their performances are shown in the SPIE paper presented into these proceedings[6]. In summary, there are two fibers, one that is used on the target (the science fiber, or fiber#2), and another one that is used on a "source-less" region (the sky fiber, or fiber#1) for sky correction when the two-fiber mode is used (see more details about observing modes in Section 2.3). The fibers transmittance is higher than 80% at wavelengths redder than 750nm, and their focal ratio degradation (FRD) numbers are < 14%.

The fibers were installed into the OHANA conduit (Figure 2) running between the CFHT dome to the Gemini one on 24 April, 2014 (watch the movie of the fiber installation at https://www.youtube.com/watch?v=Th-AW60puf8). The FRD numbers measured before and after the installation are comparable, indicating that the installation was successful. Both ends were carefully handled to prevent any twist and curl along the way. On the CFHT end, the fiber was routed for half a dozen meters down to the spectrograph. On the Gemini end, about 70m of fiber were routed across the first floor, then up the telescope pier to the telescope on the fifth floor (see Figure 3).

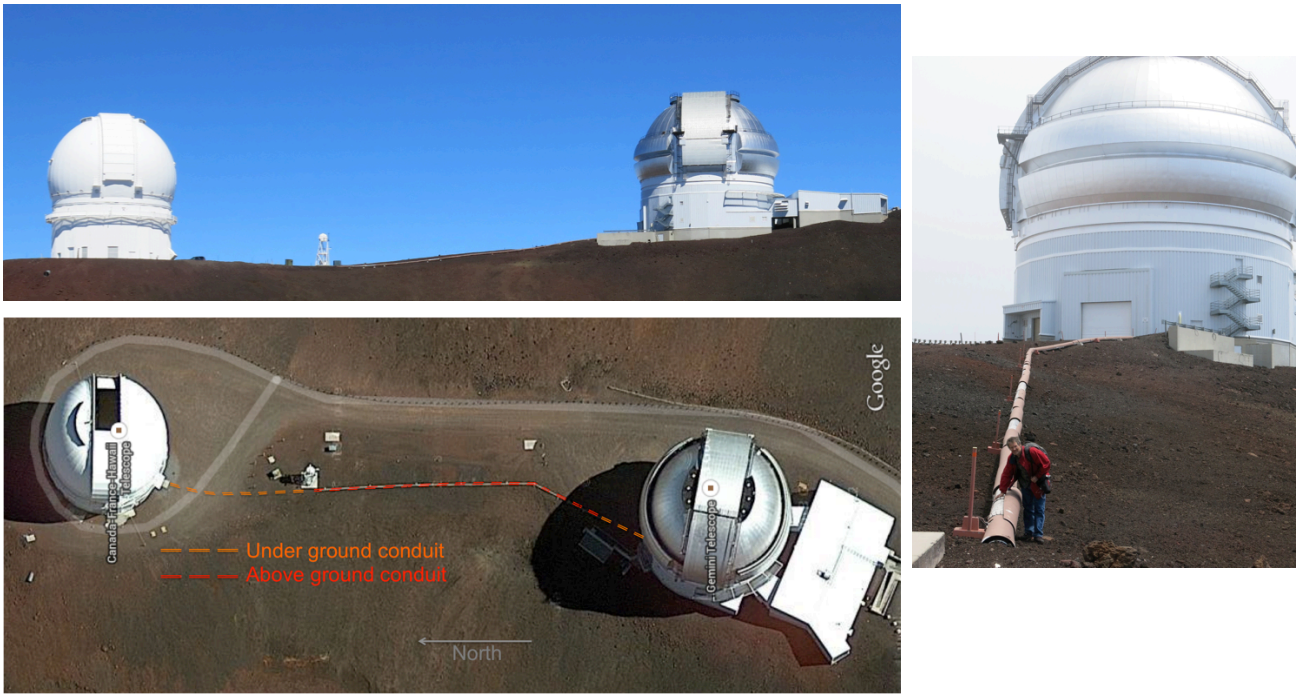

Figure 2. Different perspectives showing the OHANA conduit between CFHT and Gemini.

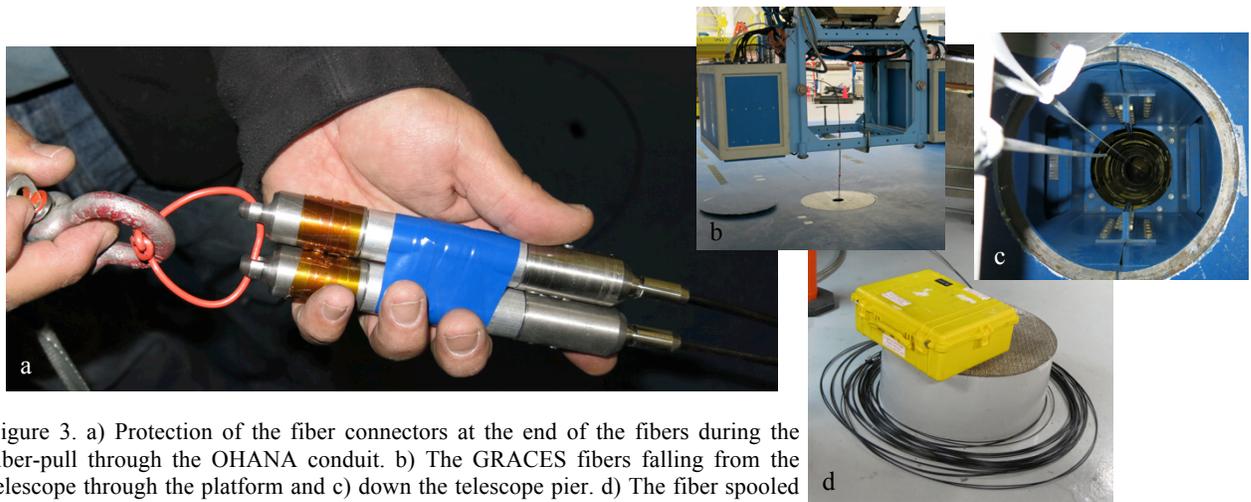

Figure 3. a) Protection of the fiber connectors at the end of the fibers during the fiber-pull through the OHANA conduit. b) The GRACES fibers falling from the telescope through the platform and c) down the telescope pier. d) The fiber spooled and stored down the pier when GRACES is not used.

## 2.3 The receiver unit

The receiver unit is the only GRACES component with moving parts controlled by a Galil 4183 servo controller. Installed onto a bridge inside the ESPaDOnS spectrograph (see Figure 4a), it contains a bench holding the optics, an image slicer, a dekker blocking "unwanted" light from the slicer, a shutter and a pickoff mirror sending the light to the spectrograph. The bench can rotate to switch between two slicing modes. One mode slices the image of the two fibers into two parts each (see Figure 4d). That mode is used when the sky is observed simultaneously with the target and delivers the smallest resolution power offered by GRACES. That mode can be called the "two-slice" mode (this name describes how many times the fiber images are sliced), but can also be referred to as the "two-fiber" mode (as the two fibers are used) or the "star+sky" mode (the most descriptive name for astronomers, as it describes better what kind of data it provides). The other mode slices the image of the fiber#2 (or science fiber) in four (see Figure 4e). That mode is called the "four-slice" mode, or "one-fiber" mode, or "star only" mode. All the stages can move back and forward between the two positions repeatedly without loosing any of the slicing performance in the two modes. It suffers from some stray light pollution, but that can be fixed with a better enclosure. Note that the original ESPaDOnS cannot be used when the receiver unit is installed into the spectrograph, as the latter is too big. For the experimental phase, we used a temporary set-up to block room light.

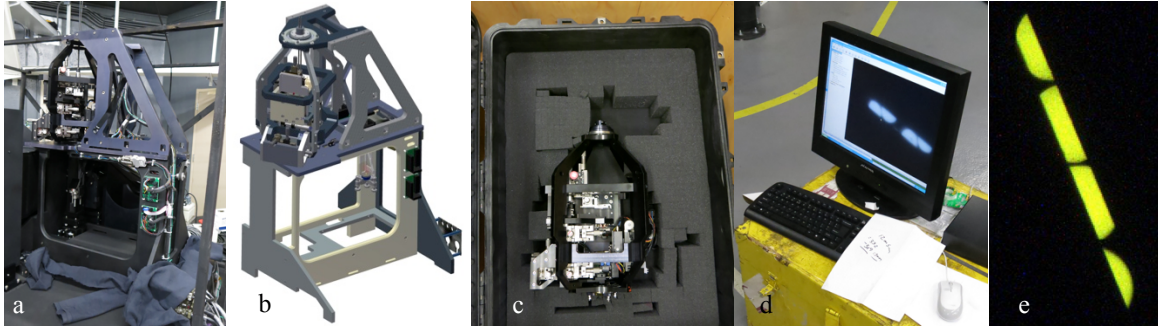

Figure 4. a) The receiver unit installed into ESPaDOnS. b) CAD model of the receiver unit installed onto its bridge. c) The receiver unit stored into its box as when GRACES is not used. d) and e) Sliced image of the fiber(s) in the two- and the four-slice modes, respectively.

## 2.4 The spectrograph and the detector

A complete description of the ESPaDOnS spectrograph is available in a previous SPIE paper[1]. If GRACES has its own fibers, image slicer and dekker, it uses all the rest of ESPaDOnS' optics. In the experimental phase of GRACES, it is not yet possible to control ESPaDOnS' parts remotely from Gemini, except for the detector. When GRACES observations are taken, the observer can determine the exposure time, the type of observation (Bias, Dark, Flat or Object) and the read mode (Fast, Normal, or Slow). The Fast read mode is used for some calibrations and bright targets, and the Slow one for the faintest targets. The Table 1 gives the readout time, the read noise and the gain for each read mode.

Table 1. ESPaDOnS' detector read modes.

| Read mode | Read time (s) | Read noise ($e^-$) | Gain ($e^-$ per ADU) |
|---|---|---|---|
| Fast | 32 | 4.7 | 1.6 |
| Normal | 38 | 4.2 | 1.3 |
| Slow | 60 | 2.9 | 1.2 |

## 3. GRACES OPERATION

GRACES operation demands lots of coordination between the CFHT and the Gemini observatories. If GRACES components can all be controlled from the Gemini control room, none of the ESPaDOnS' components can be move from anywhere but CFHT. Therefore, many tasks need to be scheduled considering both telescopes schedules and availabilities. Thanks to the very collaborative spirit of all the teams involved, this could be done quite smoothly during GRACES experimental phase. Here are some details about the different aspects of GRACES operation.

### 3.1 Components control

Communication to GRACES is done via Unix sockets and HTTP. GRACES could run on any Linux workstation in Gemini, but for security reasons, it is limited to run from a single Linux machine. The current implementation of the software is a proof of concept that lacks some of the features found in a full facility instrument. Even then, the software can control all the instrument mechanisms, retrieve data and status information, which is the minimum required to carry out observations at night. The GRACES mechanisms/devices that can be controlled from Gemini are the bench rotation, the image slicer and the dekker position. There are two commands used to take an exposure, i.e. "start exposure" (observe) and "abort exposure". No pause, resume or stop commands are provided by ESPaDOnS.

### 3.2 During daytime

The GRACES focus needs to be checked and tuned daily. The reason for this is that the temperature variations inside of the ESPaDOnS spectrograph are of the order of 1°C within a day. This gets even worst when people have been walking into the CFHT Coudé room during the day; this is why it is ideal to wait a minimum of three days between when work is done in the Coudé room and when the first observations on-sky are taken. To get the best focus value, a series of thorium-argon (ThAr) lamp spectra (see Section 4.1) is taken with the ESPaDOnS collimating mirror moved at different positions. The ThAr spectra are inspected to find the collimating mirror position giving the narrowest spectral lines, hence the best focus value. The mirror is moved from CFHT, while the light comes from the Gemini Facility Calibration Unit (GCAL - http://www.gemini.edu/sciops/instruments/gcal/?q=sciops/instruments/gcal) with the tertiary mirror pointed to GMOS+GRACES. Therefore, the focus run can be executed only when both telescopes and the respective teams are available.

ESPaDOnS is stable enough to take calibration frames only once during the day before the observing night. Three series of ten 0sec frames using the Slow, Normal and Fast readout modes are taken to correct for the bias level of ESPaDOnS' detector. Also, one series of ThAr spectra (for wavelength calibration) and one series of flat-field frames (to correct for the variations in the pixel-to-pixel sensitivity) in the two- and four-slice modes are taken (see Section 4.3).

When dark frames are requested, they are observed in the morning, using the same read mode and the same exposure time as the science frame that need to be corrected for dark current.

### 3.3 Acquisition

Acquisition of a target takes about 10 minutes in average. This includes slewing the telescope from a previous position to the requested one, acquiring the guide star and moving the target over the GRACES science fiber. The acquisition sequence is done using GMOS in imaging mode. This implies that for the whole sequence, the GRACES cassette is out of the beam, and the GMOS filter wheels can be used. The acquisition procedure consists on placing the target image on the GMOS detector at a position in pixel that corresponds to where the science fiber is moving once the GRACES cassette is placed back into the beam. This pixel position is nicknamed (and will be referred to in this paper as) the sweetspot. Once the telescope pointing at the requested position and the guide star is acquired, the sequence to put the target over the sweetspot is the following:

1. Applying an instrument offset of $p = 23.6$ arcsec and $q = -9.5$ arcsec to place the target within ±0.5 arcsec from the desired position.

2. Taking an image using a 1x1 pixel binning and at 1800x1100 pixels sub-raster centered on the GMOS CCDs (see Figure 1d). The readout time is ~13sec when the fast read/low gain mode is used, i.e. most of the time.

3. Measuring the centroid of the target image on the GMOS image.

4. If the target is within 0.1 arcsec from the sweetspot, moving to the next step. Otherwise, applying the offset to improve the target position and going back to step 2. The offset accuracy provided by the OIWFS is ~0.1 arcsec when the offset is smaller than 10 arcsec.

5. Placing the GRACES cassette into the beam, and setting the wavelength at which the system is guiding (to optimize flux at that given wavelength).

Since the GMOS filters introduce distortions to the field image, the sweetspot depends on which filter is used for the acquisition. The filter is chosen based on the target's magnitude and color. For fainter targets (R>10mag), we use broad-band filters and for brighter target, we use narrow-band filters to avoid saturation in the acquisition image. The GMOS Integration Time Calculator can be used to decide which filter, with which exposure time and which readmode is the more appropriate. Table 2 gives the pixel position for the sweetspot obtained in the GMOS filters used for GRACES.

Table 2. Pixel position (sweetspot) as a function of GMOS filter.

| GMOS filter | X (pixel #) | Y (pixel #) |
|---|---|---|
| g | 1218.45 | 433.09 |
| r | 1215.87 | 435.33 |
| i | 1218.12 | 436.86 |
| HeII | 1218.45 | 433.13 |
| HeIIC | 1218.97 | 433.04 |
| Hα | 1218.42 | 433.20 |
| HαC | 1218.83 | 433.07 |

To determine the sweetspot the first time, a bright star was acquired into one of the fiducial hole on the injection module mask (see Figure 1d). Since the distance is known between the holes and the position were the target is aligned with the fibers, a first series of gross offsets were applied. Once completed, we used the exposuremeter in ESPaDOnS. Since the night was photometric, we applied offsets in $p$ and $q$ to the target and measured the flux going through the science fiber. The results are shown in Figure 5. We see that within a 0.1 arcsec radius, the lightloss is smaller than 5%. Once the flux through the fiber was peaked and we knew the star was well centered on the fiber, we moved the GRACES cassette out of the beam and took images in different GMOS filters.

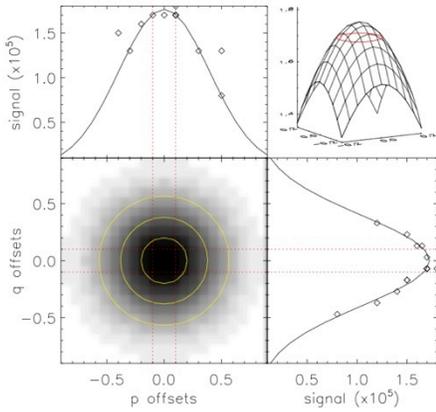

Figure 5. This figure shows the changes in the signal measured with the ESPaDOnS' exposuremeter as a function of $p$ and $q$ offsets. This result can be used to reconstruct the size of the fiber on sky, and to determine how GRACES performances are dependent on the target centering on the fiber.

### 3.4 Observation

Assuming the science target is centered on the detector sweetspot for a given filter, the GRACES cassette can be placed in position on GMOS, so the fiber gets in front of the beam. In the commissioning phase, the observations first were carried out from the CFHT remote control room (May 5-6, 2014), and then later from the Gemini remote observing room (May 13-16, 2014) using dedicated software written specifically for this phase (see Section 3.1).

The observations, from the perspective of the Gemini user, consist of choosing the two- or four-slice mode, readout mode, and exposure time. After each science exposure, the data is transferred to a Gemini server, to add proper header keywords to the fits files.

### 3.5 Guiding, flexures and light-losses

Since the GRACES cassette is installed inside GMOS, its On-Instrument Wave Front Sensor (OIWFS - http://www.gemini.edu/sciops/instruments/gmos/?q=node/10377) is used for guiding when using GRACES. A model of the OIWFS flexure is used to keep the target in position. As shown in Section 2.1, GMOS+OIWFS flexures cause a total accuracy in the acquisition of 0.07 arcsec 95% of the time (after corrections using flexure models), with a maximum uncertainty in positioning of 0.12 arcsec, but only at elevations lower than 30°. This is mostly within the tolerances of that can be derived from Figure 5 where the effect of the target centering on GRACES performances is shown.

GRACES suffers slit-losses due to atmospheric differential refraction in a similar manner GMOS can be with a wide long-slit (http://www.gemini.edu/sciops/instruments/gmos/itc-sensitivity-and-overheads/atmospheric-differential-refraction). It is therefore possible to optimize the guiding for a given wavelength using a central wavelength value.

## 4. DATA FORMAT AND EXTRACTION

### 4.1 Calibrations

The minimum required (and recommended) calibration frames are:

- 5 (20) flat-field (quartz halogen lamp) exposures;

- 1 (3) arc (Th-Ar lamp) exposure;

- 1 (3) bias exposure.

The flat-field images and ThAr arc lamps were taken using the light coming from GCAL through the fiber. In order to optimize the flux in the blue portion of the spectra without saturating the red side, we used the GMOS balance filter. Typical exposures for both flat-fields and arc lamps were 150 seconds.

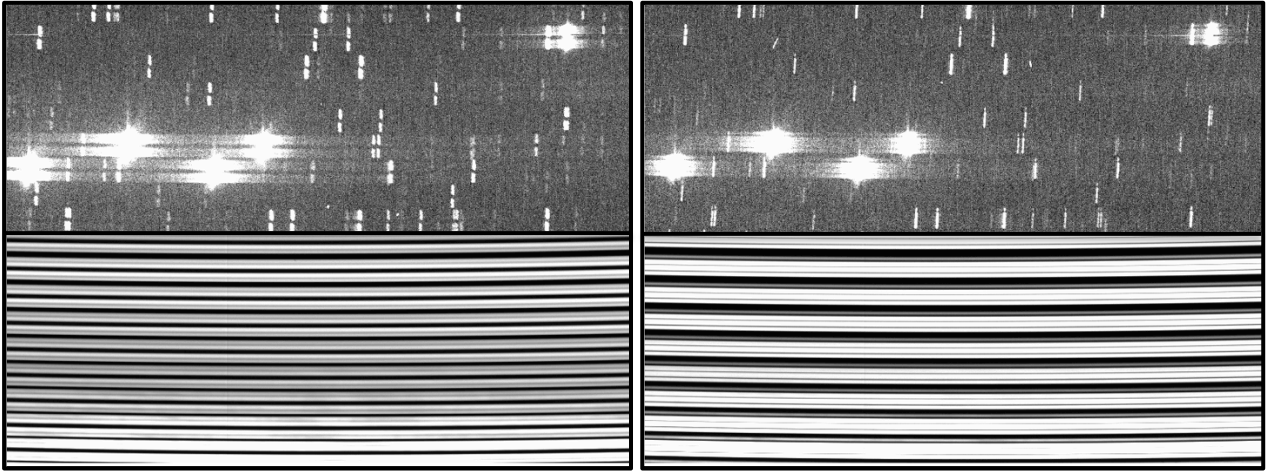

Figure 6. Sample of ThAr spectra (up) and flat field images (down) obtained in the two- (left) and four-slice modes (right). Note how clearly the sliced images can be seen on the ThAr emission lines.

### 4.2 Raw spectra

Figure 7 shows the raw 2D spectrum of GRACES first light, when the A3 star HIP57258 ($V$=9.00 mag) was observed on May 6, 2014. We see that is covers the optical band from ~405 nm to ~1.03 μm (i.e. from order 58 to order 22), however the spectrum gets very dim bluer than 420 nm. The spectrum is continuous (gap-less) until 922.5 nm, and small 1-2 nm gaps appear between the last 3 orders.

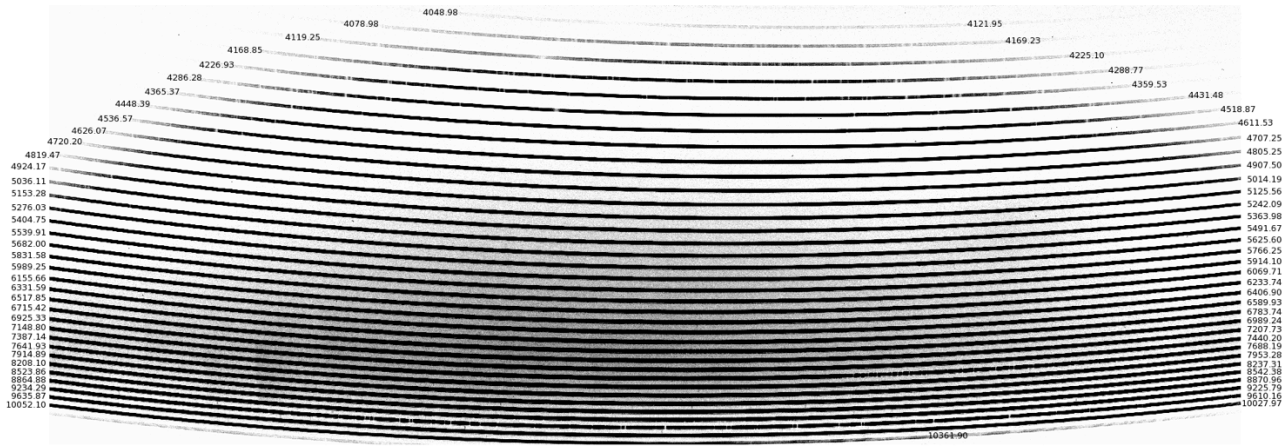

Figure 7. Raw 2D spectrum of the A3 star HIP 57258, also the GRACES first light!

### 4.3 Data reduction

For the GRACES data reduction, two different approaches were used: (i) standard IRAF[1] tasks, using the *echelle* package; and (ii) the OPERA pipeline[7], modified to process GRACES data files. Both results were comparable, and we will only describe the steps of the reduction sing OPERA in this section, as the IRAF reduction was essentially done following the standard steps from Daryl Willmarth's cookbook[8]. Moreover, OPERA is more appropriate for GRACES data, as it provides a better spectral resolution and an optimal extraction.

The reduction with OPERA starts by producing the lists of calibration and object raw files that will be used in the reduction. Then OPERA executes two steps: (1) Calibration and (2) Reduction. Each of these steps is summarized as follow.

**1. Calibration** - executes the following 6 steps:

a. *Master combining*: median combines several calibration images of the same type (bias, flat, or arc) into a single master calibration frame.

b. *Gain and Noise*: measure CCD gain and noise using on a set of flat-field and bias exposures. As an example, for data obtained on 2014-05-06 the measured gain was 1.68±0.02 e⁻/ADU and the noise was 5.51 e⁻, with a bias level of 490 ADU in Fast readout mode. Nominal gain and noise values for ESPaDOnS in Fast readout mode are 1.5 e⁻/ADU and 4.7 e⁻, respectively.

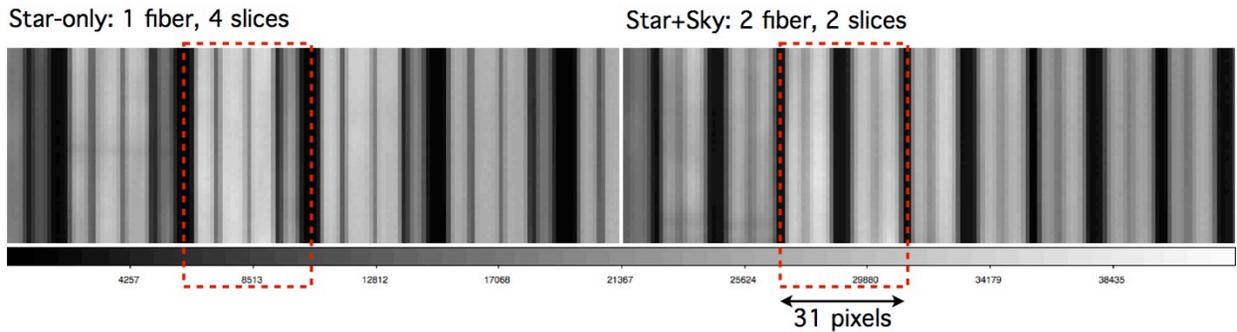

Figure 8. Left panel shows a section of a flat-field frame in Star-Only (4-slice) mode, showing a few orders in the far red. Right panel shows the same section of a flat-field frame in Star+Sky (2-slice) mode.

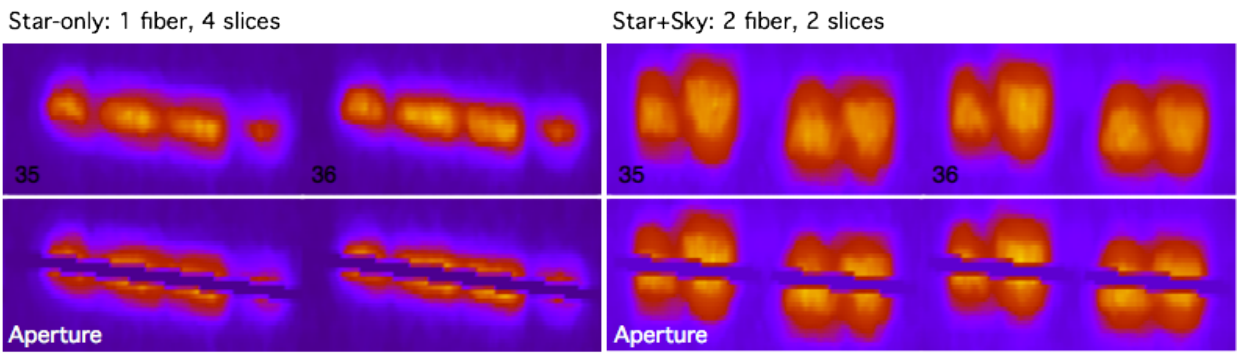

Figure 9. Top panels show measurements of the instrument profile for orders 35 and 36, where left panel is in Star-Only mode and right panel is Star+Sky mode. Bottom panels show the same measurements with aperture sub-pixels masked out.

---

[1] IRAF (http://iraf.noao.edu) is distributed by the National Optical Astronomy Observatories, which are operated by the Association of Universities for Research in Astronomy, Inc., under cooperative agreement with the National Science Foundation.

c. *Geometry*: detect and enumerate spectral orders, then calculate a polynomial function that models the center path of each order on the detector. This step was particularly challenging for GRACES because the orders start with an order separation of about 65 pixels (bluest) and it decreases down to 30 pixels for reddest orders. The spatial profile also spans about 30 pixels, which makes it difficult to distinguish adjacent orders as their separation gets smaller. For Star+Sky mode it becomes even more critical since the separation between adjacent orders is smaller than the separation between fibers (see Figure 8). This confuses the automatic identification of red orders. OPERA handles well this situation by making use of an empirical model for order separations and by matching order positions using a cross-correlation with the spatial profile.

d. *Instrument Profile*: this step performs measurements of a two-dimensional oversampled instrument profile as a function of image coordinates. GRACES presents very interesting profiles given by the irregular shape of the pseudo-slit, which is produced by the slicer. Figure 9 presents measurements of the instrument profile of GRACES for orders 35 and 36 (central wavelength at 648 nm and 630 nm, respectively) for both instrument modes as indicated in the figure.

e. *Aperture*: this step performs measurements of the tilt angle of a rectangular aperture for extraction. It uses the instrument profile to measure the tilt angle that maximizes the flux fraction inside the aperture. The calibrated aperture is aligned with the monochromatic image of the pseudo-slit, allowing unbiased flux measurements of each spectral element. Bottom panels in Figure 9 show the instrument profiles with the measured extraction apertures masked out. The average tilt measured for Star-only mode is −2.64±0.07 degrees and for Star+Sky mode is -1.63±0.04 degrees.

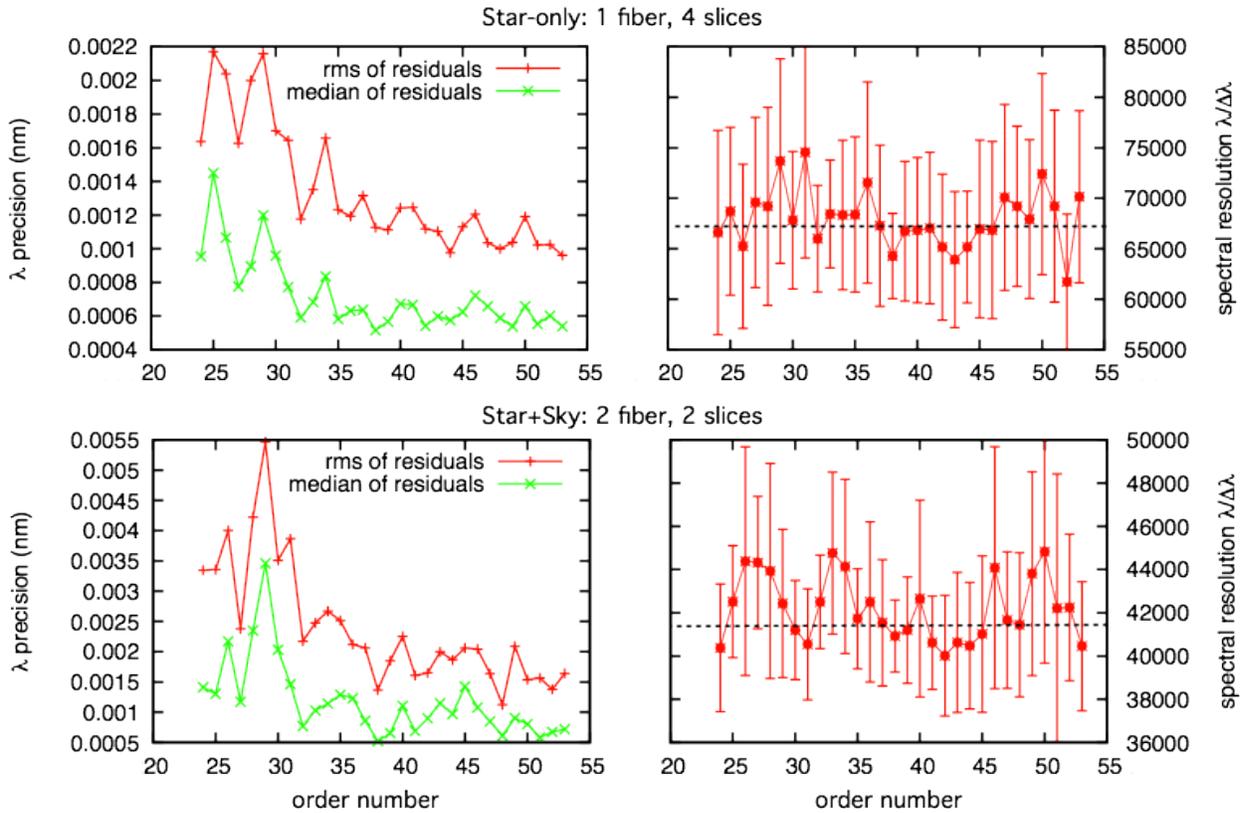

Figure 10. Left panels show measurements of the RMS (in red) and median (in green) residual wavelength of matched spectral lines, where top panel is Star-Only mode and bottom panel is Star+Sky mode. Right panels show measurements of spectral resolution where the dashed line is the average of all orders.

f. *Wavelength calibration*: this step performs a standard pixel-to-wavelength calibration, where it first detects spectral lines from a ThAr comparison extracted spectrum and compares these lines with known atlas data to create a calibration polynomial solution for each order. Left panels in Figure 10 show the RMS and median residual between central wavelength measured for each line and the wavelength given in the atlas. This represents an estimate for the wavelength precision in nm. Right panels in Figure 10 show the spectral resolution, where the mean value for Star-only mode is R=65768±406 and for Star+Sky mode is R=41161±209.

## 2. Reduction

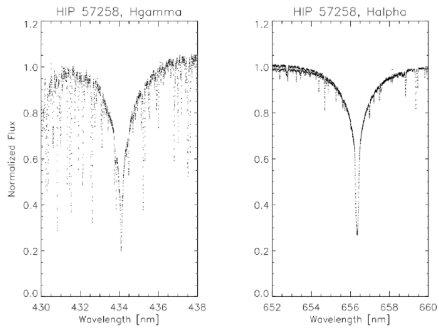

a. *Extraction*: this is the main step of OPERA reduction, where it extracts the flux spectra of object frames using the calibration quantities measured in the previous steps. It applies the Optimal Extraction algorithm of K. Horne (1986)[9] adapted for tilted apertures.

b. *Flat-fielding* (fringing correction): the relatively high levels of electronic fringing in the red part of GRACES spectrum (lower orders) can be successfully corrected by dividing the object spectrum by a flat-field extracted spectrum, where both should be reduced using the same calibrations.

Figure 11. Extracted spectrum around Hα and Hγ of HIP57258.

## 5. GRACES PERFORMANCE

### 5.1 Resolution

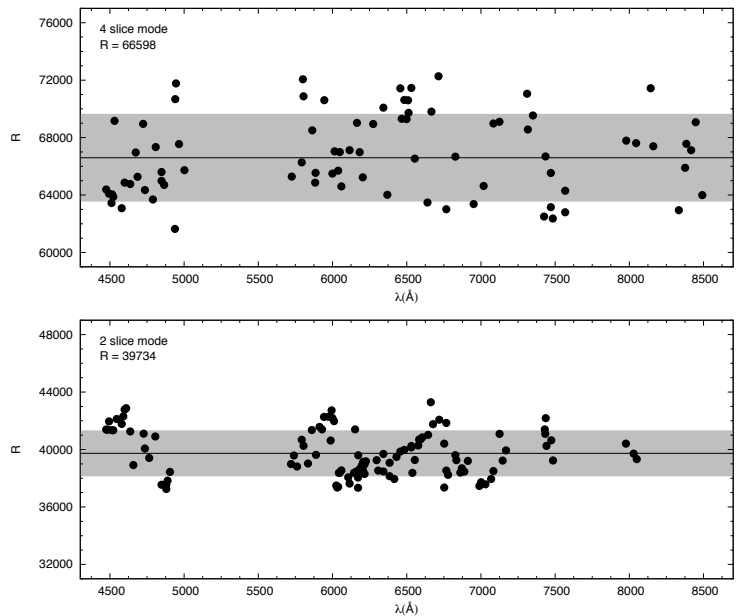

Figure 12. Resolution power derived from a ThAr raw spectrum as a function of wavelength for the two- (lower) and the four-slice mode (upper).

The spectral resolution was calculated based on 110 lines (two-slice mode) and 82 lines (four-slice mode) of the ThAr lamp. The average pixel size for the two-slice mode is 2.88 pixels and 1.74 pixels for the four-slice mode. Figure 12 shows the resolution as a function of the wavelength, from 4500 to 8500 angstroms. The solid lines are the averages for each mode, and the shaded areas represent ±1 sigma from the mean values.

For the two-slice mode, the average resolution is about R~40,000, while for the 4-slice mode, the GRACES resolution reaches an average of over R~66,000. These values are 20% higher than the predictions obtained from simulations (R=33,000 and 55,000). They are also *slightly higher* than the values derived with OPERA, but this is because the reduction software derives the resolution power from extracted spectra, and not from the raw spectrum like here.

## 5.2 Sensitivity and throughput

The star Feige66 was observed early during GRACES commissioning to assess the sensitivity and throughput. It is an OB subdwarf star with an effective temperature of 34,500K and a logarithmic surface gravity *log(g)* = 5.83, that is commonly used as spectrophotometric standard[10]. Its *Vmag* is 10.54.

The sensitivity for GRACES reflects the magnitude of an object that would provide a signal-to-noise ratio of 1 for an hour of integration time ($S/N_{1hr}$). To calculate the signal-to-noise ratio, the signal (*S*) is determined from a 600sec observation of Feige 66. That observation was obtained under pristine weather conditions, with photometric sky and a seeing patch close to 0.3 arcsec. The air mass at the time of the observation was 1.13, and the spectrum was therefore corrected to an air mass of 1 using the extinction values (mag per air mass) from Béland (1988)[11]. The spectrum was also binned into the resolution element, i.e. 1.7 pix and 2.9 pix for the four- and the two-slice mode, respectively, so the final result is presented as a function of resolution element.

The noise is determined from a subtracted pair of 1200sec frames on sky, observed shortly after the Feige66 spectrum. With a subtracted pair of sky frames, on an area of sky per order generally free of sky lines and artifacts, the noise (*N*) is determined by multiplying the average standard deviation in the background counts by the square root of the aperture area used to extract the signal.

$$S/N_{1hr} = \frac{S \cdot (3600sec/T_{Feige66})}{N \cdot (3600sec/(2 \times T_{sky}))^{1/2}}, \qquad (1)$$

where $T_{Feige66}$ and $T_{sky}$ are the exposure time of the Feige66 spectrum and the sky, respectively. To find the sensitivity, the signal-to-noise ratio ($S/N_{1hr}$) is converted to a magnitude and added to the Feige66 spectrum calibrated in magnitude ($M_{Feige66}$) available from the catalog of Massey[10]:

$$Sensitivity = 2.5 \times log(S/N_{1hr}) + M_{Feige66}. \qquad (2)$$

The result for GRACES sensitivity is plotted in Figure14. This results means that GRACES can deliver a signal-to-noise ratio of 1 in 1 hour for a target as faint as 21.9 mag.

The throughput measurement is based on the same observation that was used for sensitivity, and the same value of *S* is extracted. This value is then compared with the Feige66 spectrum calibrated to the flux of photons hitting the primary mirror. This spectrum is obtained from the spectrum observed with the Hubble space telescope ($S_{Feige66}$) available in the literature[12] to which we apply a coefficient to correct for the atmosphere extinction. In summary:

$$Throughput = \frac{S}{S_t}, \qquad (3)$$

where

$$S_t = S_{Feige66} \cdot 10^{(-0.4 \cdot \mu \cdot A_\lambda)}, \qquad (4)$$

where μ is the air mass and $A_\lambda$ is the extinction coefficient from Béland (1988)[11]. The result is presented in Figure 15.

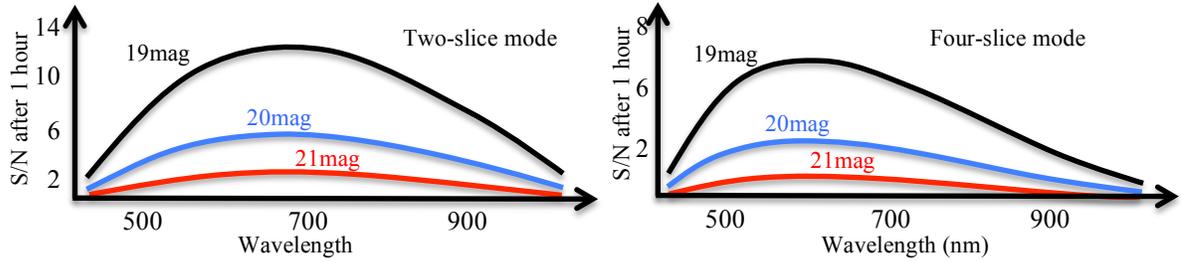

Figure 13. S/N reached for a flat spectrum of different magnitudes after a 1h exposure in the two- (left) and the four-slice mode (right).

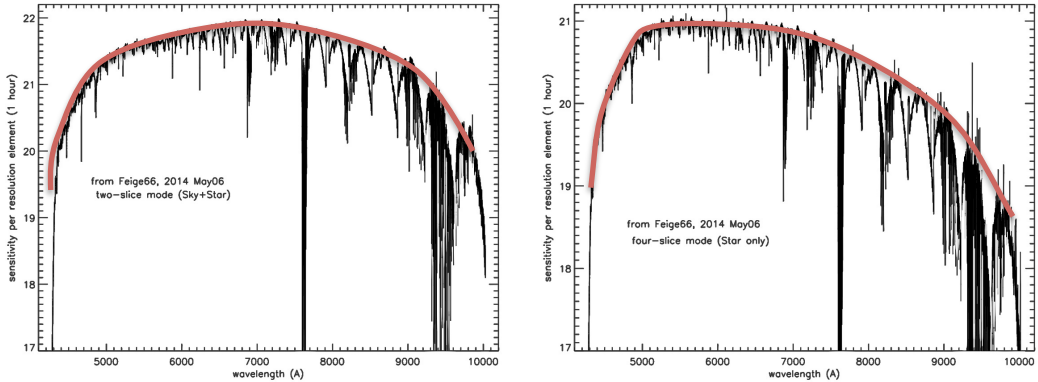

Figure 14. GRACES sensitivity, defined as the magnitude of a target giving a S/N~1 after a 1h exposure for the two- (left) and the four-slice mode (right).

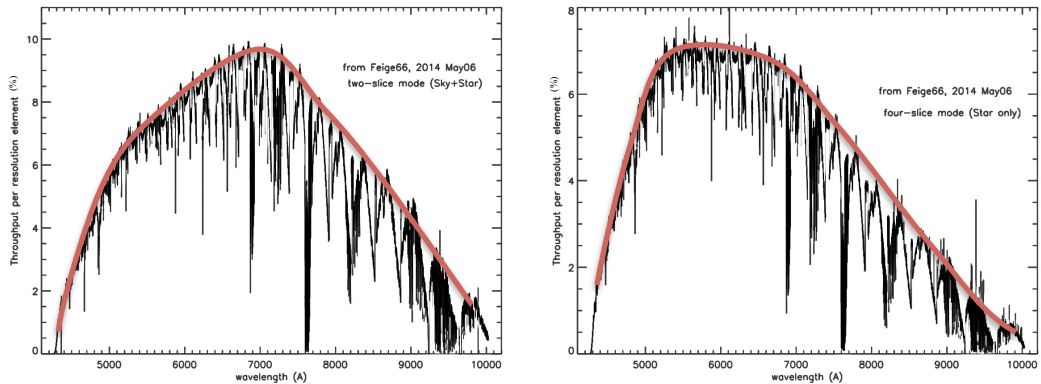

Figure 15. GRACES throughput (corrected for the atmosphere extinction).

### 5.3 Comparison with HIRES

The first obvious comparison for GRACES is HIRES on KeckI. The mirrors are of comparable size, and both observatories are on top of Mauna Kea. Moreover, HIRES offers resolution powers that are comparable to GRACES. Comparing the performances of two different instruments on two different telescopes can be misleading. The final result depends strongly on many assumptions, and on the definition of which parameters are compared. Here, we present the results of what we consider the fairest comparison of our observations with comparable spectra of Feige 66 observed with HIRES.

The HIRES slit giving a resolution that matches that of GRACES in two-slice mode is C5 (1.15 arcsec). We have downloaded the raw spectrum of an observation using the C5 slit from the Keck archive KOA (http://www2.keck.hawaii.edu/koa/public/koa.php). The spectrum was obtained on April 21, 2008, it has a wavelength range of 5320Å-10180Å, and was observed using a gg475 blocking filter. The fits filename is HI.20080421.18828.fits and was observed for the program U066Hr (PI: Koo).

It is not possible to know from the archive webpage or the file header if the night was photometric or not, but it is among the spectra with the highest S/N within comparable wavelength ranges in the archive. We can estimate the seeing at the time of the observation using the width of the spectra's trace. The binning used for the two spectra was 2 pixels in the spatial axis, giving a pixel scale of 0.24 arcsec, and we estimate a seeing (FWHM) of 0.83 arcsec. This corresponds to a slit-loss of ~10%. Losses due to seeing were virtually inexistent with GRACES observations.

After scaling the HIRES spectrum to match the GRACES (better) weather conditions and resolution power, and binning all the spectra to one resolution element (2.88 pixels for GRACES and 3.2 for HIRES), we compare the *S/N* obtained 1 hour on Feige 66 with both spectrographs in Figure 16. Note that the result for HIRES matches the result obtained with the Exposure Time Calculator (http://etc.ucolick.org/web_s2n/hires), when the parameters are set as shown in Figure 17.

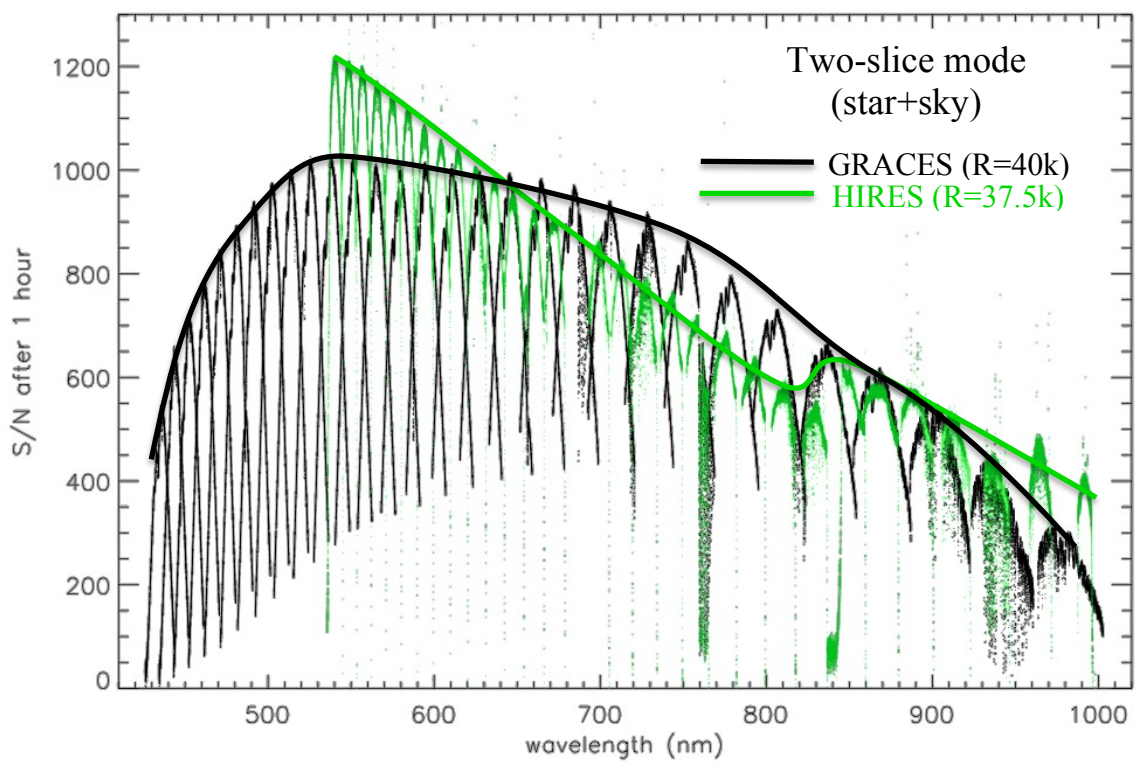

Figure 16. Comparison of the GRACES spectrum of Feige 66 observed in the two-slice mode, and the HIRES spectrum leveled to match GRACES observations weather conditions and resolution.

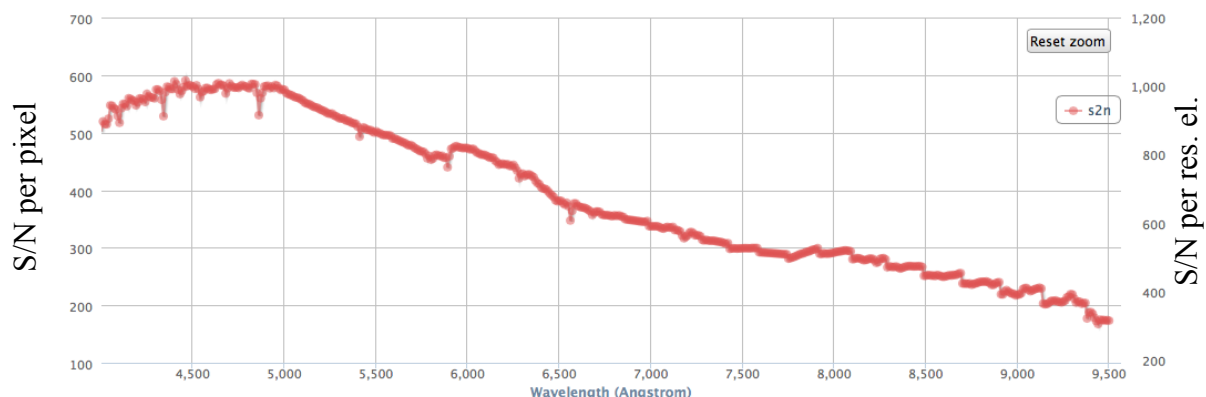

Figure 17. Result giving the predicted *S/N* of a 1 hour spectrum of Feige 66 using HIRES with the C5 slit.

We also compare the GRACES four-slice mode to an archive HIRES spectrum using the slit B2 (0.57 arcsec) giving a resolution of 67 000. We have downloaded the raw spectrum obtained on June 5, 2006. It has a wavelength range of 4570Å-9140Å, and was observed using a kv408 blocking filter. The fits filename is HI.20060605.33359.fits and was observed for the program H288Hr (PI: Kleyna). We estimate that the spectrum was observed with a seeing of 0.61 arcsec. This corresponds to a slit-loss of ~43%. However, even if the observations were taken under a seeing of 0.4 arcsec, like the GRACES spectrum was, it would have led to a 9% slit-loss. Therefore, we only apply an effective correction of 31% to the HIRES spectrum. The comparison of the spectra is done the same way we did for the two-slice mode, and the result is presented in Figure 18.

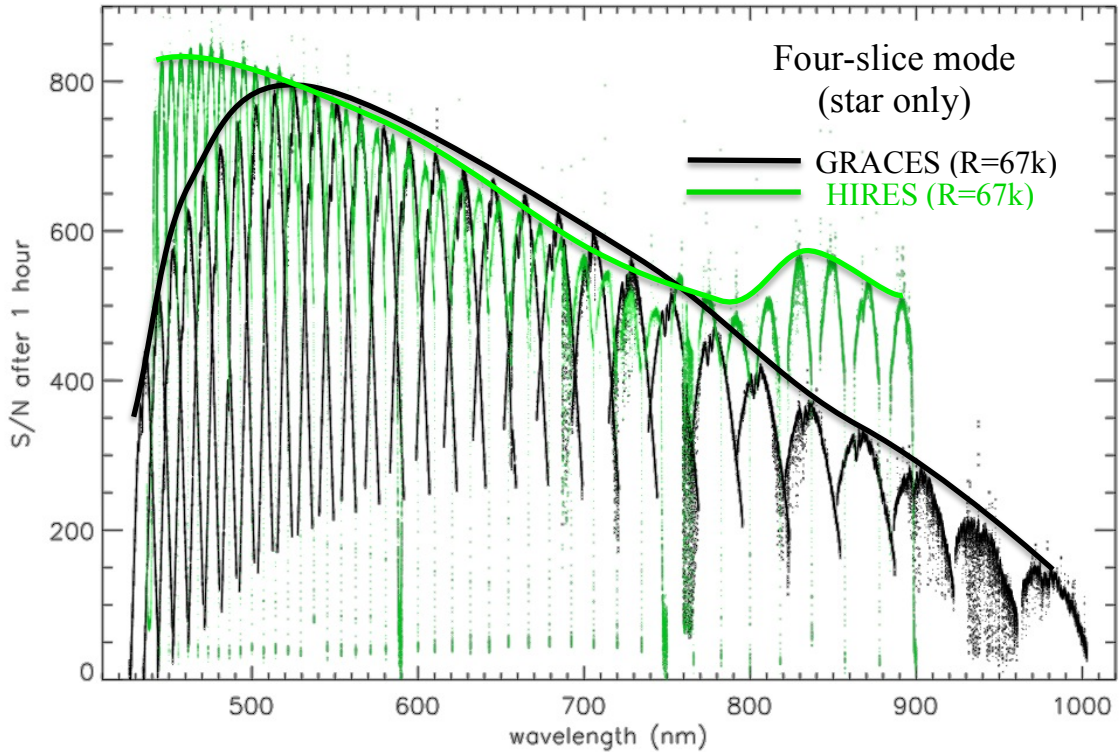

Figure 18. Comparison of the GRACES spectrum of Feige 66 observed in the four-slice mode, and the HIRES spectrum leveled to match GRACES observations weather conditions and resolution.

In conclusions, both spectrographs are pretty comparable. This is an amazing result, considering that it is a 270m-long fiber that links the mirror to the spectrograph. It is as if it was not even there! This is a huge advancement in fiber technology that opens many new opportunities in astronomy. In terms of new capability at Gemini, GRACES successfully fulfills its role.

**5.4 Other performances**

Spectra of faint targets, of spectrophotometric standards observed at different air masses and of short-period binaries were also observed to assess the on-sky GRACES sensitivity and its accuracy in radial velocity, as well as verifying how it is affected by atmospheric differential refraction. All those results will be hosted on the GRACES webpage (http://www.gemini.edu/sciops/?q=node/12131).

## 6.  CONCLUSIONS

The GRACES experimental phase was a **complete success**! Not only were we able to send light from the telescope through the fibers, but also the resulting high-resolution spectrograph is giving performances comparable to other instruments on 6-10m telescopes. In other words, the fibers are so great that we almost forget they are there. GRACES is not possible without the great collaboration between the CFHT, Gemini, NRC and ESPaDOnS teams. Minor modifications of some optics (nothing related to the fibers) are now needed to improve GRACES performances by up to 20%. Also, different teams have already expressed the interest to build a polarimetric device or an iodine cell for GRACES.

*Acknowledgments:*
We are very grateful to all the numerous teams who made this challenging project successful, and more specifically to the day crews at both CFHT and Gemini for their kind contribution and for making GRACES installation and operation possible.